\documentclass[num-refs]{wiley-article}
\usepackage{siunitx}
\usepackage{geometry}
\usepackage{longtable}
\usepackage{multirow}
\usepackage{booktabs}
\usepackage{verbatim}
\usepackage{totpages}
\usepackage[utf8]{inputenc}
\usepackage{textgreek}

\papertype{Original Article}

\paperfield{Journal Section}

\title
{How GenAI Mentor Configurations Shape Early Collaborative Dynamics: A Classroom Comparison of Individual and Shared Agents}

\author[1]{Siyu Zha}
\author[1]{Weijing Liu}
\author[1]{Fei Qin}
\author[2]{Jie Cao}
\author[3]{Yanjin Wang}
\author[4]{Yujia Liu}
\author[5]{Kaiyi Zhang}
\author[6\authfn{1}]{Jiangtao Gong}
\author[1]{Yingqing Xu}

\affil[1]{Tsinghua University, Beijing, China}
\affil[2]{University of North Carolina at Chapel Hill, USA}
\affil[3]{City University of Hong Kong, Hong Kong SAR, China}
\affil[4]{Carnegie Mellon University, USA}
\affil[5]{Xi'an Jiaotong University, China}
\affil[6]{Shanghai Jiao Tong University, China}

\corraddress{JiangTao Gong, Shanghai Jiao Tong University, Shanghai, China}
\corremail{gongjiangtao2@gmail.com}

\runningauthor{Author One et al.}

\begin{document}

\begin{frontmatter}
\maketitle

\begin{abstract}
 

Generative artificial intelligence (GenAI) is increasingly embedded in computer-supported collaborative learning (CSCL), yet little empirical research has unpacked how different configurations of AI participation reshape collaborative processes. This study investigates how GenAI configuration shapes collaborative regulation in authentic classroom settings. Two eighth-grade classes engaged in small-group creative problem-solving under two conditions: a shared-AI configuration, in which each group interacted with a single AI mentor, and an individual-AI configuration, in which each student accessed a personal AI instance. Using multi-layer discourse coding combined with lag sequential analysis (LSA) and ordered network analysis (ONA), we examined interaction distribution, AI–student coupling, shared regulation processes, and teacher orchestration. Results reveal distinct regulatory dynamics across configurations. Shared AI access promoted convergence-oriented collaboration, with stronger alignment of shared regulatory states and more coordinated group-level reasoning. In contrast, individual AI access distributed support across learners, producing more exploratory and evaluative cycles but also more fragmented interaction patterns, accompanied by increased teacher intervention to manage divergence. These findings suggest that AI configuration functions as a structural design variable that reorganizes the regulatory ecology of classroom collaboration.

\keywords{Generative AI, CSCL, Collaborative Regulation, AI Configuration}
\end{abstract}
\end{frontmatter}

\section*{Practitioner notes}
\noindent\textbf{What is already known about this topic}
\begin{itemize}
\item Generative AI is increasingly integrated into classroom learning, yet most studies focus on individual use rather than collaborative, group-based learning contexts.
\item Prior work in Computer-Supported Collaborative Learning (CSCL) demonstrates that interaction structures and tool configurations significantly shape socially shared regulation of learning (SSRL).
\item The introduction of AI tools into group work presents new orchestration demands for teachers, particularly when students shift attention toward AI-generated outputs.
\end{itemize}

\noindent\textbf{What this paper adds}
\begin{itemize}
\item Provides classroom-based empirical evidence that AI access configuration (shared vs. individual) is associated with distinct interaction distributions, AI–student coupling patterns, and regulatory dynamics in CSCL.
\item Shows that shared AI access is linked to convergence-oriented collaboration and stronger group-level alignment, whereas individual AI access is associated with more distributed, exploratory, and repair-oriented regulatory cycles.
\item Conceptualizes AI configuration as a structural design variable that reorganizes the regulatory ecology of collaborative learning rather than merely affecting individual performance.
\end{itemize}

\noindent\textbf{Implications for practice and/or policy}
\begin{itemize}
\item Shared AI access was associated with stronger group-level alignment and coordinated regulation, suggesting it may be better suited for tasks requiring collective sense-making.
\item Individual AI access was associated with more distributed and parallel engagement patterns, indicating potential benefits for exploratory phases, though additional synthesis support may be required.
\item AI configuration should be treated as a pedagogical design choice rather than a technical decision. Schools and instructional designers may benefit from aligning configuration models with different task phases and ensure that teacher have appropriate support to sustain collaborative regulation.
\end{itemize}

\clearpage
\section{Introduction}

Generative artificial intelligence (GenAI) is rapidly entering collaborative learning environments \cite{kovari2025systematic,feng2025group,wei2025effects}. A growing body of research highlights its pedagogical potential, demonstrating benefits for student engagement, feedback provision, and scalable classroom integration \cite{zhuang2025enhancing,belkina2025implementing,cukurova2025interplay}. Beyond serving as an instructional aid, GenAI increasingly participates in collaborative dialogue, contributing ideas, explanations, and evaluative feedback that shape collective meaning-making \cite{guoa2025student,zheng2025generative}. As AI shifts from a supportive tool to a collaborative partner, an important yet underexplored question emerges: how does the structure of AI access reorganize collaborative interaction and regulation in authentic classrooms? While existing studies primarily examine AI’s functional affordances, they seldom consider access configuration itself as a structural factor shaping participation and coordination in group-based Computer-supported Collaborative Learning (CSCL) settings\cite{liu2025integrating,ogunleye2024systematic,perifanou2025collaborative,chen2025analyzing}.

Decades of CSCL research suggest that collaborative processes are highly sensitive to tool structure and orchestration design. Variations in how resources are distributed within groups can alter interactional patterns, cognitive coupling, and socially shared regulation of learning (SSRL)\cite{jarvela2015enhancing,malmberg2015promoting}. From this perspective, AI access configuration may not merely affect how frequently students consult AI, but how regulation unfolds across individuals and groups. Additionally, \cite{salomon1992effects} emphasized that the design and implementation of new technologies in educational settings must be examined within the broader classroom context. Yet empirical evidence from authentic classroom settings remains limited, particularly regarding how shared versus individual AI access shapes regulatory dynamics over time.

Addressing this gap, the present study compares two Grade 8 classrooms engaged in identical creative problem-solving(CPS)\cite{treffinger1995creative} tasks using the same GenAI system, differing only in access configuration. In condition A, each group interacted with a shared AI instance. In condition B, each student accessed an individual instance. By holding AI's functionality constant and varying only the access structure, we investigate how configuration influences interaction distribution, AI–student coupling, regulatory transitions, and teacher orchestration in authentic CSCL settings. Accordingly, this study addresses four research questions:

\begin{itemize}
    \item RQ1: How does AI configuration influence the distribution of interaction events among students, the AI, and the teacher?
    \item RQ2: How does AI configurations shape the coupling between AI functions and students’ behavioral and cognitive responses?
    \item RQ3: How does AI configuration reshape patterns of socially shared regulation within student groups?
\end{itemize}

By examining AI configuration within a real classroom CSCL context, this study moves beyond outcome comparisons and provides process-level insights into how structural design choices shape collaborative regulation and orchestration. These findings contribute to a more ecologically grounded understanding of AI-supported collaborative learning.

\section{Theoretical Background}

The rapid integration of GenAI into collaborative classrooms introduces new forms of interaction, coordination, and regulation. Rather than functioning solely as instructional aids, GenAI systems increasingly participate in dialogue, propose ideas, generate explanations, and provide evaluative feedback\cite{sharples2023towards,teng2024chatgpt}. These contributions position AI as an active component within collaborative learning systems. To conceptualize how different AI access configurations may reshape classroom collaboration, this study integrates perspectives from distributed cognition, CSCL, and socially SSRL.

\subsection{Human--AI Collaboration as Distributed Cognitive Activity}

Human--AI collaboration in education can be conceptualized through the lens of distributed cognition, which views cognitive activity as emerging from the coordination of individuals, artifacts, and shared representations \cite{stahl2013theories,liu2025bricksmart}. Within CSCL environments, learning unfolds through interaction among peers, tools, and external representations rather than residing solely within individual minds. GenAI increasingly functions as a cognitive partner in this distributed system, contributing explanations, suggestions, and evaluative feedback that become integrated into collaborative dialogue \cite{yan2024promises,davis2025co}. In this sense, GenAI participation is not merely additive; it reorganizes how reasoning is distributed across human and artificial actors.

Empirical studies demonstrate that GenAI can scaffold collaborative sense-making by expanding representational resources and structuring discussion~\cite{zha2025designing}. For instance, AI-supported shared displays have been shown to facilitate collective brainstorming and comparison of ideas \cite{zhang2025ladica}, while LLM-generated dialogue maps can provide common reference points that sustain group awareness \cite{chen2025meetmap}. Multi-agent approaches further suggest that generative AI can enhance knowledge elaboration and collaborative problem-solving outcomes compared with more traditional chatbot configurations \cite{zheng2025generative}. These findings indicate that AI systems can serve as shared reference points for reasoning within CSCL settings.

At the same time, emerging evidence suggests that AI participation can also redistribute attention and regulatory responsibility within groups. Learners may increasingly orient toward AI rather than peers during problem solving \cite{feng2025group}, potentially altering patterns of coordination and shared regulation. Such shifts raise an important but still under examined question: not whether AI supports learning in general, but how different structures of AI access influences the distribution of interaction, cognitive coupling, and socially shared regulation in authentic classroom settings. Addressing this question requires moving beyond functional affordances to examine configuration-level effects on collaborative dynamics.

\subsection{GenAI Mentor Configurations and the Structuring of Interaction}

CSCL research has consistently shown that learning technologies do more than deliver information; by affording and constraining particular actions, tools structure patterns of participation, coordination, and meaning-making within collaborative activity \cite{roschelle1995construction,suthers2006technology}. From this perspective, introducing GenAI into collaborative problem solving does not merely add a resource but reorganizes how attention, authority, and interpretive responsibility are distributed within the group. As AI systems increasingly generate explanations, suggestions, and evaluative feedback, they become active components of the socio-technical system.

Emerging evidence suggests that AI participation can shift the focus of interaction within groups. Learners have been observed engaging more frequently with chatbots than with peers during problem-solving processes, raising concerns about shifts in attention and influence may be redistributed \cite{feng2025group}. At the same time, when AI-generated outputs are made visible to all members, they may function as shared representational anchors that coordinate collective reasoning. Experimental studies in triadic human–human–AI settings further indicate that shared AI access can reduce over-reliance and strengthen collective accountability for interpreting AI outputs \cite{daryanto2026human}. Similarly, platforms that combine shared and individual AI panels demonstrate that visibility and distribution of AI contributions shape engagement and collaborative flow \cite{sayeed2025collaclassroom}.

Overall, these findings suggest that AI access configuration—whether support shared at the group level or individually distributed—may fundamentally structure interaction by shaping who consults the AI, how its outputs circulate, and how regulatory responsibility is allocated. However, empirical evidence examining such configuration effects at the process level in authentic classroom CSCL settings remains limited. A configuration-sensitive lens is therefore needed to understand how different AI access structures reshape participation patterns and regulatory dynamics in collaborative learning.

\subsection{Socially Shared Regulation of Learning in AI-mediated Collaboration}

Collaborative problem solving depends on groups’ ability to regulate their learning processes collectively. The SSRL framework conceptualizes regulation as a co-constructed process through which learners jointly set goals, align interpretations, monitor progress, and coordinate strategies \cite{hadwin2011self,jarvela2015enhancing}. In traditional CSCL contexts, shared artifacts and prompts can trigger negotiation, alignment, and monitoring processes.

Introducing GenAI adds a new layer to these regulatory dynamics~\cite{giannakos2025promise}. AI-generated suggestions can prompt reflection, spark negotiation, or shift the direction and momentum of groupwork \cite{law2025role}, \cite{giannakos2025promise}. In shared-AI configurations, decisions about when and how to consult the AI become collective regulatory acts \cite{han2024teams}. In individual-AI configurations, students may regulate their work in parallel through interactions with their personal AI instances. While this structure can support individual elaboration, it  potentially reducing opportunities for shared monitoring~\cite{zercher2025can}. Because SSRL concerns how regulation is distributed and coordinated across group members, it provides a critical lens for examining how AI access configuration reshapes collaborative processes over time~\cite{kim2025socially}.

These perspectives position AI configuration as a structural variable within a triadic socio-technical system, influencing participation patterns, cognitive coupling, and regulatory dynamics. Building on this framework, the present study investigates how shared and individual GenAI mentor configurations gave rise to different collaborative interaction and regulation in authentic classroom settings.

\section{Methodology}

\subsection{Research Context and Participants}

The study was conducted in a public middle school in China as part of a multi-week creative problem-solving (CPS)program. This CPS project contained several phases, including problem framing, idea exploration, solution generation, and prototype development. The present study focuses on the problem framing stage, during which students defined the problem space and began developing preliminary solution directions. This early phase was chosen because configuration differences are most likely to manifest in how groups initiate regulation processes, coordinate roles, and incorporate AI support \cite{li2022influence}.

Two Grade 8 classes (N = 38, ages 13–14) participated in the study. In each class, students were organized into seven small groups of three to four members. The two classes were comparable in prior academic achievement and demographic composition. While the school had previous experience integrating AI to support learning, it had not previously implemented AI mentors in collaborative CPS tasks. The study was conducted during regular instructional time and aligned with school-wide initiatives in AI literacy and computational thinking. Parental consent and student assent were obtained in accordance with institutional and school policies.

\subsection{Experimental Conditions: Individual-Agent vs. Shared-Agent AI Mentors}

Both classes completed the same CPS task and shared identical learning objectives, instructional materials, teacher, and classroom structure. The only difference was the configuration of the GenAI mentor, which was asssigned at the class level to prevent cross-condition contamination.

\textbf{Individual-Agent Condition.} In condition A, each student interacted with a personal GenAI mentor through a laptop or tablet. Students could pose questions, request explanations, or seek guidance independently at any time during collaboration. The design provided personalized and parallel scaffolding, allowing students to explore ideas individually before brining them to group discussions.

\textbf{Shared-Agent Condition.} In condition B, each group interacted with a single shared GenAI mentor through a centrally positioned device. Group members collaboratively negotiated which prompts to send and jointly interpreted AI's responses. This configuration emphasized collective regulation, requiring explicit coordination, turn-taking, and shared decision-making regarding tool-use.

The same teacher instructed both conditions. No explicit guidance was provided regarding when or how frequentlt students should consult the AI, allowing natural interaction pattern to emerge.

\subsection{AI Mentor System and Technical Background}

The AI mentor used in this study was adapted from an educational dialogue system previously developed by our research team. For this classroom implementation, we tailored the system to fit this CPS project theme, which focused on designing solutions for a “Smart Garden”. The AI system incorporated domain-relevant knowledge and example prompts to support students in exploring environmentally conscious design ideas, interpreting design constraints, and developing early conceptual directions.

The interface was powered by GPT 5.0 capable of sustaining multi-turn conversations. We designed a structured role prompt to positioned the AI as a supportive CPS mentor, guiding students through goal clarification, explanation, idea generation, reflective questioning, and procedural scaffolding. While the system incorporated a customized knowledge base aligned with the course theme, its responses were generated in a general conversational manner and did not rely on retrieval from external sources. All interactions with the AI were text-based, and the system recorded each prompt and response with corresponding timestamps.

To ensure student safety and age-appropriate use, the AI was configured to avoid unsupported claims, harmful suggestions, or inappropriate content. Importantly, the study did not manipulate AI's behavior across conditions: both classes used the same AI system, and only the access structure—individual or shared—differed. This design allows observed differences in interaction patterns to be attributed to the configuration rather than to differences in the model itself.

\subsection{Data Collection}

A multimodal dataset was collected to capture the dynamics of student–AI–teacher interaction. Classroom video and audio recordings documented both whole-class instruction and small-group dialogue, including students' spoken exchanges, moments of AI use, and the teacher’s instructions All group conversations were transcribed and segmented into meaning units for analysis. The screen recording of AI usage and AI–student interaction log provided a complete record of every prompt and corresponding AI response.

Additionally, we also drew on teacher observation notes, which highlighted key moments of clarification, redirection, or guidance related to tool use and group regulation. Brief interviews conducted with students after the activity offered additional insight into their perceptions of AI support, how they collaborated, and what challenges they encountered. Togther, these data sources enabled a fine-grained analysis of how collaboration unfolded in each condition.

\subsection{Multi-layer Coding Framework and Interpretive Orientation}

All interactional data were analyzed using a multi-layer coding framework designed to capture the  cognitive, behavioral, affective, and regulatory dimensions of collaboration. Drawing on the conceptual perspectives outlined earlier, this framework included codes  for socially shared regulation, student cognitive and behavioral moves, affective expressions, teacher orchestration, and the functional roles enacted by the AI mentor. Rather than treating these layers as separate variables, we conceptualized them as interdependent components of a triadic system involving students, the AI, and the teacher.

\begin{table}[ht]
\centering
\caption{Overview of coding categories used in the interaction analysis}
\begin{tabular}{p{3cm} p{3cm} p{7cm}}
\hline
\textbf{Category} & \textbf{Example Codes} & \textbf{Description} \\
\hline
Student orientation & S-S, S-AI, S-T & Social, teacher-directed, and AI-directed interaction focus \\
Student affect & A1--A3 & Emotional valence in interaction \\
Student cognition & C1--C5 & C1–C4: task-related cognitive depth; C5: non-task-related \\
Student behavior & B1--B8 & Inquiry, response, probing, revision, evaluation behaviors \\
Shared regulation (SSRL) & SR-A, SR-B, SR-C & Affective, behavioral, and cognitive shared regulation \\
AI functions & F1--F7 & Explanation, prompting, evaluation, generation, encouragement \\
Teacher regulation & TR-C, TR-B, TR-A & Cognitive, behavioral, and affective regulation moves \\
Teacher behavior & TB1, TB2 & Teacher-initiated and responsive actions \\
\hline
\end{tabular}
\label{tab:code_overview}
\end{table}

Transcripts were coded at the level of interactional turns. Four trained coders engaged in iterative calibration to ensure shared coding criteria, and disagreements were resolved through discussion. Reliability checks conducted on a randomly selected subset of the data yielded substantial to near-perfect agreement across coding categories (κ = .78–.87). Due to space constraints and to maintain readability, the main text provides only brief conceptual descriptions of the coding dimensions(see Table 1), while the full codebook with detailed operational definitions and examples is included (see Table 3 in the Appendix).

\subsection{Data Analyses}

The analytic strategy combined descriptive, temporal, and structural approaches to examine how collaborative dynamics differed across AI configurations. Analyses were conducted at the collaboration-group level, corresponding to the unit of instructional assignment and statistical inference.

\subsubsection{Descriptive and Between-Condition Comparisons}

To establish contrasts in participation patterns, we first examined the relative distribution of interaction codes, including student, teacher, and AI turns; cognitive and regulatory moves; and AI consultation patterns. Because interaction volume varied across groups under authentic classroom conditions, proportions rather than raw frequencies were calculated within each coding family. For each group, the frequency of a code was divided by the total number of coded instances within that family.

Between-condition differences were estimated as contrasts in collaboration-group mean contrasts ($\Delta = A - B$). Uncertainty was quantified using bootstrap resampling (5,000 iterations) to generate 95\% confidence intervals. Statistical significance was evaluated using permutation tests (10,000 permutations), in which condition labels were randomly reassigned across groups while preserving original group sizes. Two-sided \textit{p}-values were computed based on the null distribution of mean differences, with $\textit{p} < .05$ indicating statistical significance.

\subsubsection{Lag Sequential Analysis (LSA)}

To examine moment-to-moment interactional contingencies, Lag Sequential Analysis (LSA) was conducted on turn-level behavioral transitions. LSA identifies statistically significant sequential dependencies between coded interactional moves. Transition matrices were constructed for each condition, and adjusted residuals were computed to determine significant transitions. A threshold of $p < .01$ was applied to identify the most robust and statistically significant sequential transitions. This analysis enabled comparison of sequential interaction patterns across the shared-AI and individual-AI configurations.

\subsubsection{Ordered Network Analysis (ONA)}

To investigate structural differences in cognitive progression, Ordered Network Analysis (ONA) was employed. ONA models co-occurrence patterns among cognitive states within sliding interaction windows, enabling comparison of network structures across conditions. A moving window size of three consecutive utterances was used to capture local cognitive associations. Condition-level networks were compared by examining mean network positions and subtracted mean networks, allowing identification of relative strengths in directed cognitive connections between the two configurations.

\subsubsection{First-Order Markov Transition Analysis}

To examine how AI configuration shaped regulatory dynamics, transition patterns were modeled using a first-order Markov model (FOMM), in which the probability of a state depends on the immediately preceding state. This framework was applied to two transition types.

First, AI-function–to–SSRL transitions were estimated to capture how AI outputs triggered subsequent regulatory responses. Directed turn-level transitions were defined from an AI function (F1–F7) to the immediately following student SSRL state (SR-A1 to SR-C4). Transition probabilities were obtained by row-normalizing the transition matrices within each condition.

Second, internal SSRL transitions were modeled within student–student (S–S) segments to examine peer-level regulatory dynamics. Only temporally adjacent S–S turn pairs were included, ensuring that transitions reflected co-constructed regulatory processes. SSRL-to-SSRL transition probabilities were estimated using the same FOMM procedure.

For both transition types, group-level matrices were aggregated within each condition, and between-condition differences were calculated as $\Delta = A - B$. Difference matrices were visualized using heatmaps with fixed symmetric color scales to support consistent interpretation across figures.

\section{Findings}

\subsection{RQ1: How does AI configuration influence the distribution of interaction events among students, the AI, and the teacher?}

\begin{figure*}[h]
    \centering
    \includegraphics[width=1\linewidth]{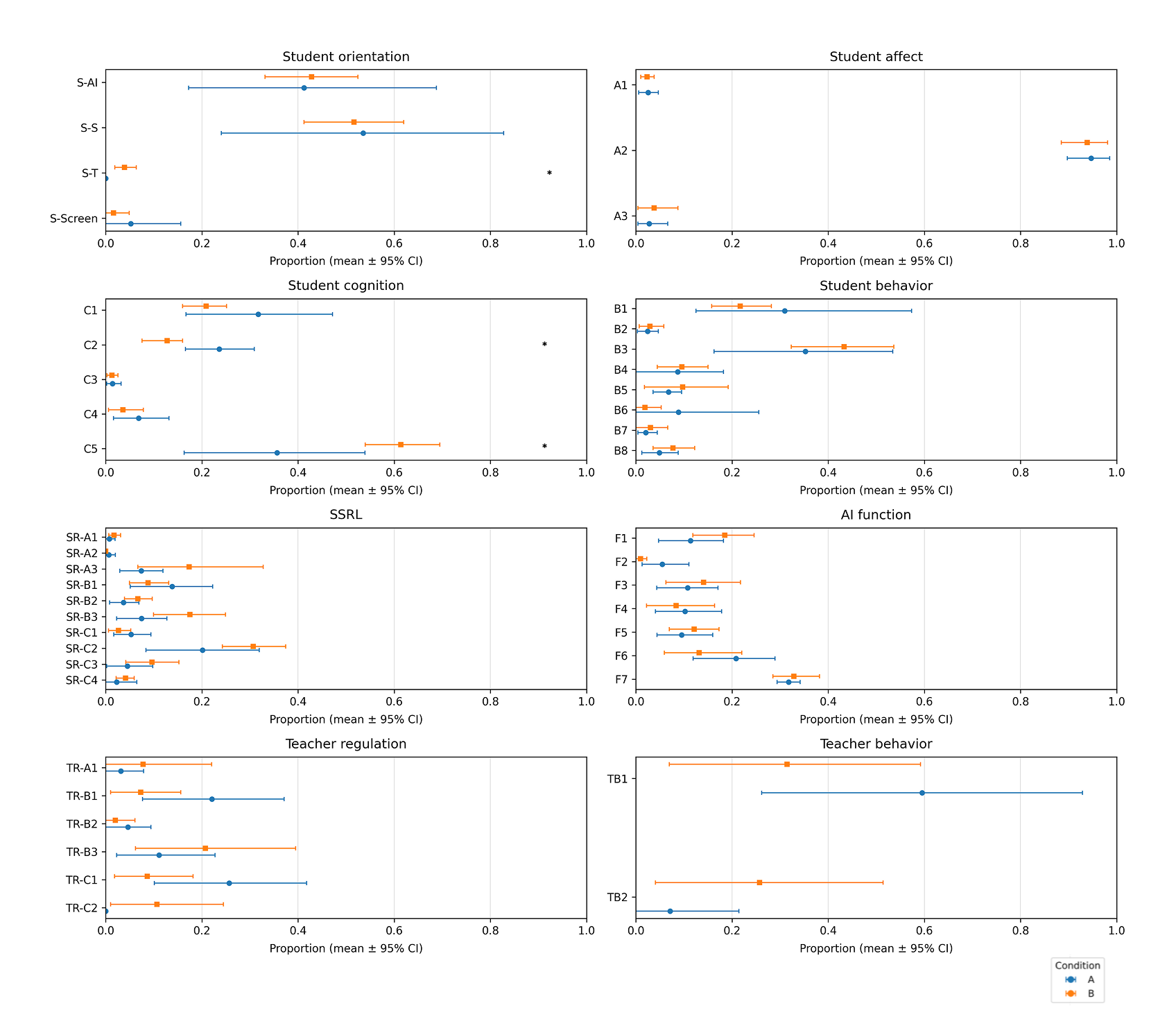}
    \caption{Mean proportions and 95\% bootstrapped confidence intervals of all interaction codes in Condition A and Condition B.}
    \label{fig:findings 1}
\end{figure*}

To address RQ1, we examined how interaction events were distributed across students, AI, and teachers under the two AI configurations. Condition B, in which students workwith individual AI agents, generated substantially more student–AI (S–AI) turns. In contrast, Condition A, which used a shared AI system, showed a higher proportion of student–student (S–S) exchanges. Moreover, student–teacher (S–T) interaction was nearly absent in Condition A but occurred more frequently in Condition B, a pattern confirmed by the significant between-group difference reported in Table 2 (\textit{p} = .005). These patterns are also reflected in the Student orientation panel of Figure 1, where the mean proportions and confidence intervals clearly distinguish the two interaction structures. Together, the results suggest that individual AI access shifted students’ attention toward the AI and the teacher, whereas the shared-AI configuration maintained a stronger focus on peer interaction.

\begin{table}[ht]
\centering
\caption{Summary of statistically significant between-group differences.}
\begin{tabular}{p{3.2cm} p{1.1cm} p{3.8cm} r r r r}
\toprule
\textbf{Category} & \textbf{Code} & \textbf{Label} & \textbf{A (\%)} & \textbf{B (\%)} & \textbf{$\Delta$ (A--B)} & \textbf{$p$} \\
\midrule
Student orientation & S-T & Student--Teacher & 0.00 & 3.90 & -3.90 & 0.005$^{**}$ \\
Student cognition & C5 & Non-task-related & 35.66 & 61.36 & -25.70 & 0.043$^{*}$ \\
Student cognition & C2 & Implementation \& analysis & 23.63 & 12.81 & +10.82 & 0.049$^{*}$\\
\bottomrule
\end{tabular}
\label{tab:summary_sig}

\vspace{0.5em}
\small
\textit{Note.} $^{*}p < .05$, $^{**}p < .01$. 
$p$ values were obtained using permutation tests.
\end{table}

Differences also appeared in the distribution of cognitive codes. As shown in Table 2, Condition B had a significantly higher proportion of C5 (non-task-related) (61.36 vs. 35.66, \textit{p} = .043), suggesting a greater incidence of off- task behavior in the individual-AI condition. In contrast, Condition A produced a significantly greater proportion of C2 (implementation \& analysis) (23.63 vs. 12.81, \textit{p} = .049). These tendencies are reflected in the Student cognition panel of Figure 1, where Condition A students contributed more analytically oriented turns, while Condition B students produced more uncategorized or low-elaboration moves. Other behavioral and affective categories were largely comparable across conditions (see Table 4 in the Appendix ).

Taken together, these findings demonstrate that the two AI configurations shaped students’ participation structures in distinct ways. The shared-AI setup fostered more peer-coordinated engagement, as reflected in higher levels of student–student interaction and more analytically oriented contributions. In contrast, the individual-AI setup led to a more externally oriented interaction pattern, characterized by more reliance on AI and teacher support and a larger proportion of off-task actions. These descriptive differences provide the structural foundation for the interactional dynamics examined in the subsequent research questions.

\subsection{RQ2: How does AI configurations shape the coupling between AI functional moves and students’ behavioral and cognitive responses?}

To address RQ2, we examined how interaction structures among students, the AI, and the teacher differed under the two AI configurations. The LSA transition networks (Fig.2) and ONA subtracted model (Fig.3) together reveal clear contrasts in how sense-making processes unfolded in the shared-AI (Condition A) and individual-AI (Condition B) settings.

\begin{figure*}[h]
    \centering
    \includegraphics[width=1\linewidth]{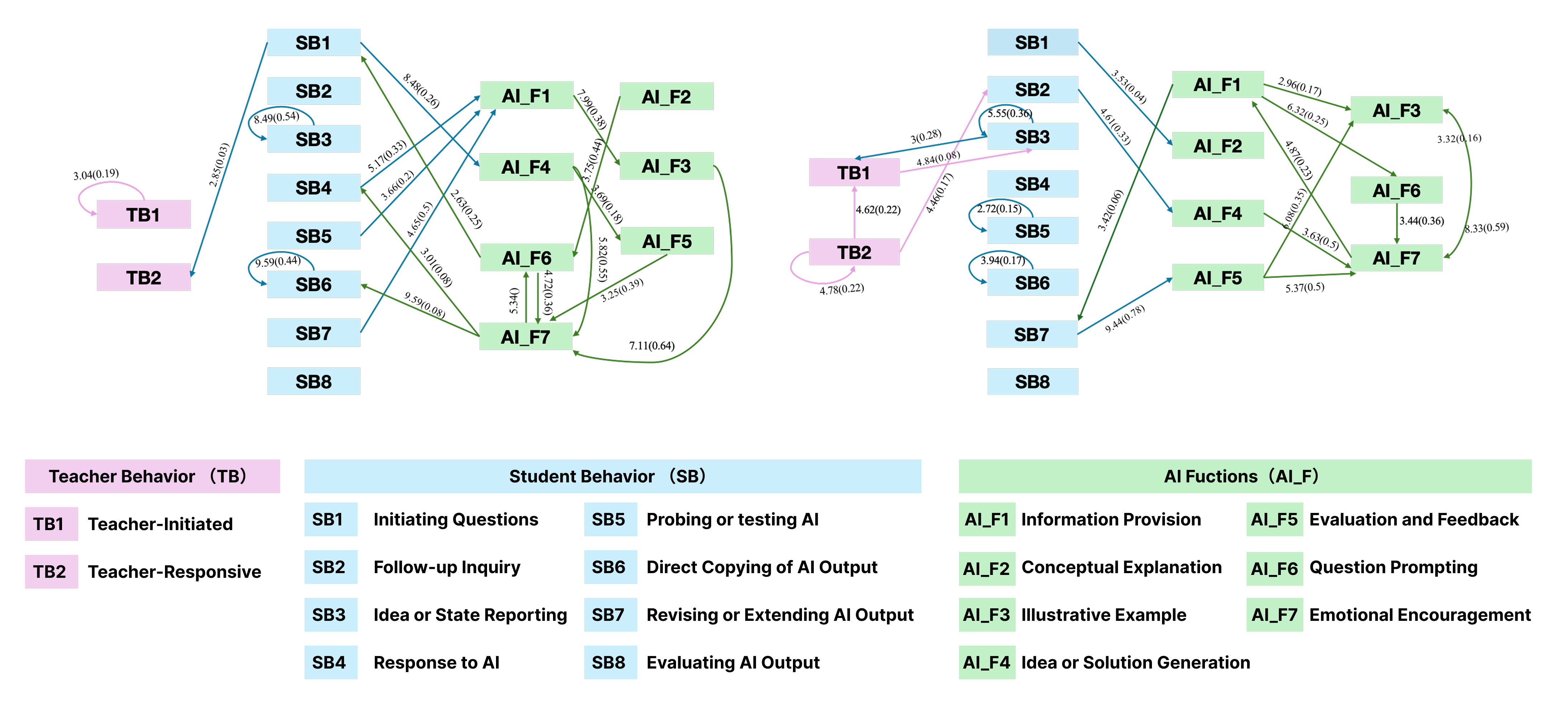}
    \caption{Lag Sequential Analysis of Interaction Patterns Under Two AI Configurations (shared-AI (Condition A) on left, individual-AI (Condition B) on right)}
    \label{fig:findings 1}
\end{figure*}

In Condition A (shared-AI), several student behaviors—including SB4 (responding to AI), SB5 (probing AI output), and SB7 (revising or extending AI output)—more frequently transitioned to AI\_F1 (information provision). This convergence suggests that AI-generated informational content functioned as a central reference point within the group. Rather than forming tightly closed loops, these transitions indicate recurrent pathways in which students repeatedly directed responses and elaborations toward the shared AI, reinforcing its role as a common informational anchor. Additional transitions from AI\_F7 (emotional encouragement) to SB4 and SB6 (direct copying of AI output), as well as from AI\_F6 (question prompting) to SB1 (initiating questions), further suggest that AI outputs often elicited immediate student uptake and follow-up questioning. Structurally, the network in Condition A reflects a more centralized configuration, where multiple student moves converge on shared AI functions and AI prompts redistribute attention back to the group.

In Condition B (individual-AI), the transition structure appears more sequential and dyadic. SB1 (initiating questions) frequently led to AI\_F2 (conceptual explanation), while SB2 (follow-up inquiry) often transitioned to AI\_F4 (idea generation). Moreover, transitions such as AI\_F1 (information provision) → SB7 (revising or extending AI output) → AI\_F5 (evaluation and feedback) formed multi-step chains within the student–AI interaction. Compared with Condition A’s convergent structure, Condition B exhibits more extended student–AI–student–AI sequences, suggesting iterative revision–evaluation cycles occurring within individual pairings. Overall, the interactional pattern in Condition B is characterized less by convergence toward a shared anchor and more by localized, sequential exchanges between individual students and their respective AI agents.

Teacher involvement also differed across the two configurations, influencing how student–AI coupling unfolded. In Condition A, teacher activity mainly took the form of TB1, TR-B1/B2 (whole-class guidance), while student-initiated help-seeking was almost absent (Table 2). The shared AI acted as a common reference point during group work, reducing the need for teacher intervention. In Condition B, both teacher-initiated scaffolding and student requests for assistance occurred more frequently. Teachers often stepped in to address divergent outputs produced by different AI agents and to help students interpret feedback from their individual AI. This pattern indicates that the individual-AI configuration redistributed orchestration demand onto the the teacher, particularly in the form of individualized support.

Cognitive transitions identified in the ONA results further highlight this distinction. In Condition A, stronger links were obserbed among C1 (comprehension), C2 (analysis/implementation), C4 (coordination), and C5(non-task-related), with notable self-transitions on C1 and C4. These patterns reflect sustained efforts to build shared understanding, clarify ideas, and coordinate next steps after receiving AI outputs. Condition B showed a relative concentration in C3 (evaluation), indicating that interactions with individual AI agents more often prompt students to make evaluative comments or judgments, but were less likely to follow up with subsequent collaborative engagement in the group.

\begin{figure*}[h]
    \centering
    \includegraphics[width=0.5\linewidth]{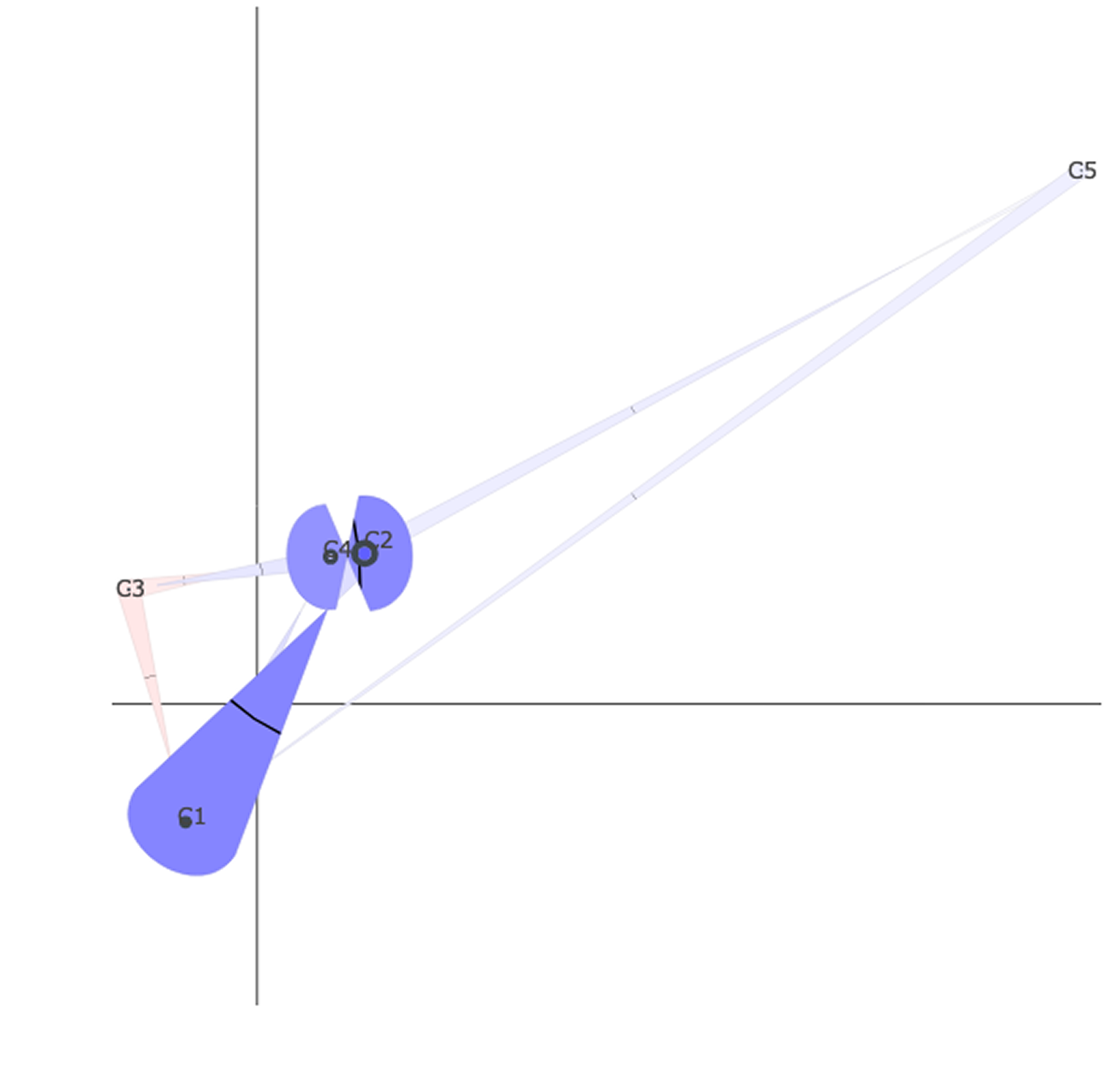}
    \caption{Subtracted ONA network (B–A) highlighting relative interaction patterns, with red indicating stronger transitions in Condition B and blue indicating stronger transitions in Condition A.}
    \label{fig:findings 1}
\end{figure*}

These findings show that shared AI access fostered more coordinated group-level engagement with students building on one another's ideas and followed up collectively. In contrast, individual AI access led to more personalized but more fragmented exchanges, often requiring greater teacher involvement to address diverging directions. In this way, the configuration of AI access shaped not only how students interacted with AI, but also how cognitive progressions formed and the degree of  teacher regulation during collaborative learning.

\subsection{RQ3: How does AI configuration reshape patterns of socially shared regulation within student groups?}

To investigate how AI functions influenced students’ regulatory responses, we compared the transitions from AI functional moves (F1–F7) to students’ shared regulation behaviors (SSRL). The difference heatmap (see Fig.~4) highlights clear contrasts between the shared-AI (Condition A) and individual-AI (Condition B) in how AI outputs prompted students' subsequent cognitive, affective, and behavioral regulation.

In the shared-AI condition, generative and evaluative AI functions were more likely to elicit group-level affective regulation. Transitions such as F4 → SR-A3 (idea generation → learning climate regulation) and F5 → SR-A3 (evaluation/feedback → learning climate regulation) were both stronger in Condition A. When the shared AI offered creative suggestions or evaluative feedback, students often responded by reinforcing group morale, strengthing motivation and aligning their positions  (''That's a wonderful suggestion! Let’s include a temperature sensor in our project design''). In these moments, students treated the AI output as a shared reference point, using it to regulate the group’s affective climate. 

\begin{figure*}[h]
    \centering
    \includegraphics[width=0.6\linewidth]{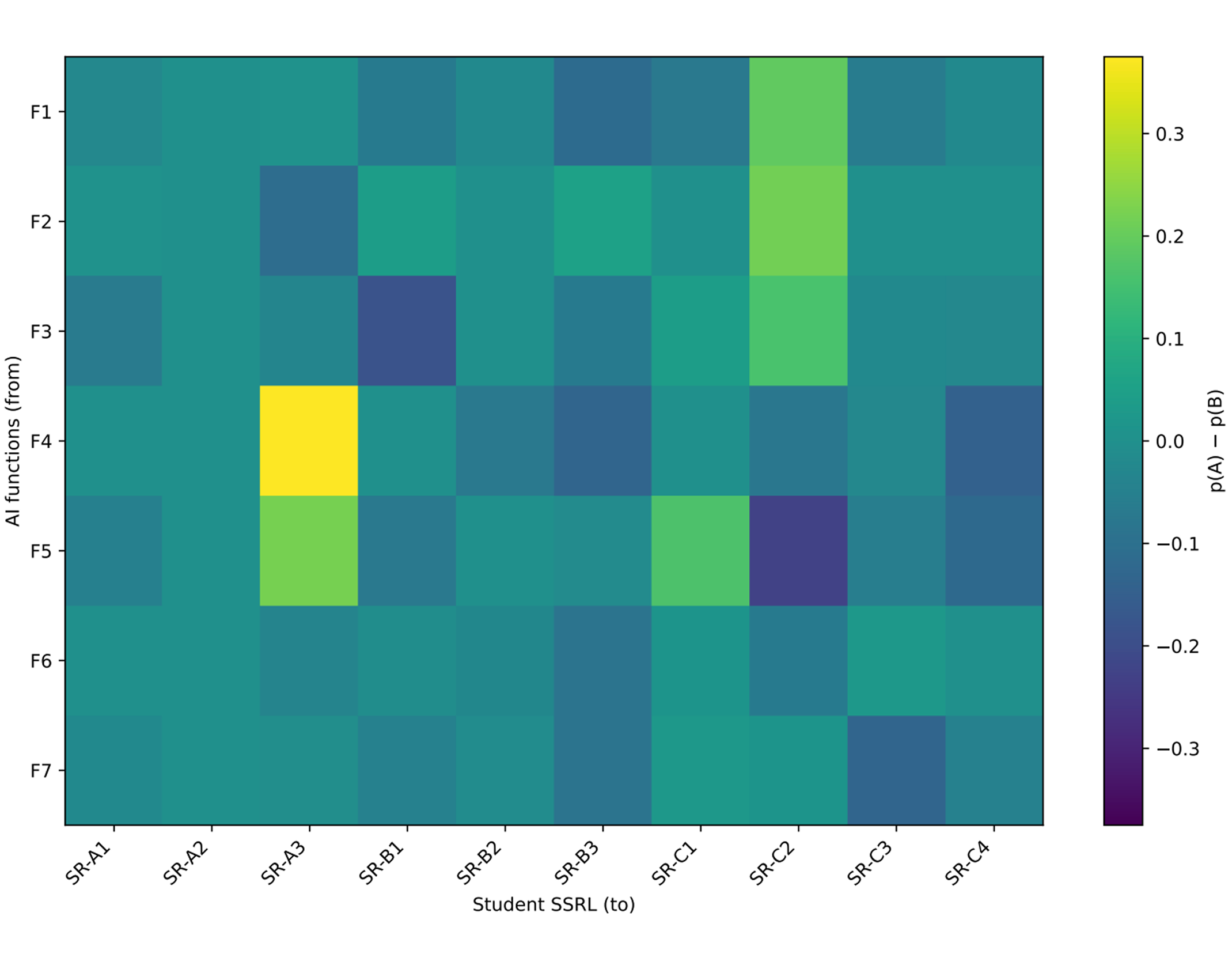}
    \caption{Heatmap of differential transitions from AI functions to student shared regulation behaviors, comparing the shared-AI (Condition A) and individual-AI (Condition B) configurations.}
    \label{fig:findings 1}
\end{figure*}

These patterns suggest that, in the shared-AI configuration, AI outputs functioned as a collective affective anchor for the group. Generative ideas and evaluative feedback were not used merely to inform individual reasoning; rather, they became socially circulated resources that shaped how the group sustained enthusiasm, reinforced confidence, and maintained engagement. In addition, the transition F2 → SR-C2 (concept explanation → collective understanding) was more pronounced in Condition A, where AI explanations frequently prompted immediate group-level alignment and conceptual clarification. Such sequences indicate coordinated socially shared regulation of learning (SSRL), in which AI contributions were collectively interpreted and integrated into the group’s shared understanding.

In contrast, the individual-AI condition showed stronger patterns of AI-driven repair and adjustment, rather than regulation initiated by the group. The transition F5 → SR-C2 (evaluation/feedback → collective understanding) occurred more frequently in Condition B. However, this “collective understanding” was often reactive, emerging when students confronted divergent outputs from their individual AI agents (“How should we frame our question?”, “What question did you enter?”"I don't know about you guys. I just typed and paste from AI."). Instead of building understanding together as in Condition A, students in Condition B often work to reconcile inconsistencies across AI responses. Similarly, the transitions from F3 → SR-B1 (example guidance → coordinated action) was more common in Condition B. In these cases, students often moved into actions by following AI's examples: not because of prior group agreement, but because AI provided a direct procedural next step. This pattern indicates an AI-led, rather than group-driven, regulatory sequence.

These patterns illustrate two distinct forms of AI–student coupling during collaborative learning. In Condition A, the AI functions acted as shared cognitive anchors, supporting students in co-constructing meaning, aligning perspectives, and maintaining a positive group climate. Regulation in this setting was largely group-centered and proactive. In Condition B, by contrast, AI functions more often prompted localized or fragmented regulation, driven by the need to coordinate across multiple individualized AI outputs.. Regulation therefore emerged as a reaction to AI inconsistencies rather than a proactive group strategy. These differences suggest that AI configuration fundamentally shapes the underlying motives and rationale of students' regulatory behavior, whether regulation through group agreement or arises in response to AI-generated divergence.


While the above findings illustrate how specific AI functions triggered distinct regulatory responses, they do not yet reveal whether these responses stabilized into sustained regulatory structures within groups. To examine this, to compare how students coordinated their work under the two AI configurations, we examined the transition patterns among SSRL states. The difference heatmap (Fig.~5) reveals notable contrasts between the shared-AI (Condition A) and individual-AI (Condition B) conditions in the cohesion, sequencing, and synchronization of shared regulation.

\begin{figure*}[h]
    \centering
    \includegraphics[width=0.6\linewidth]{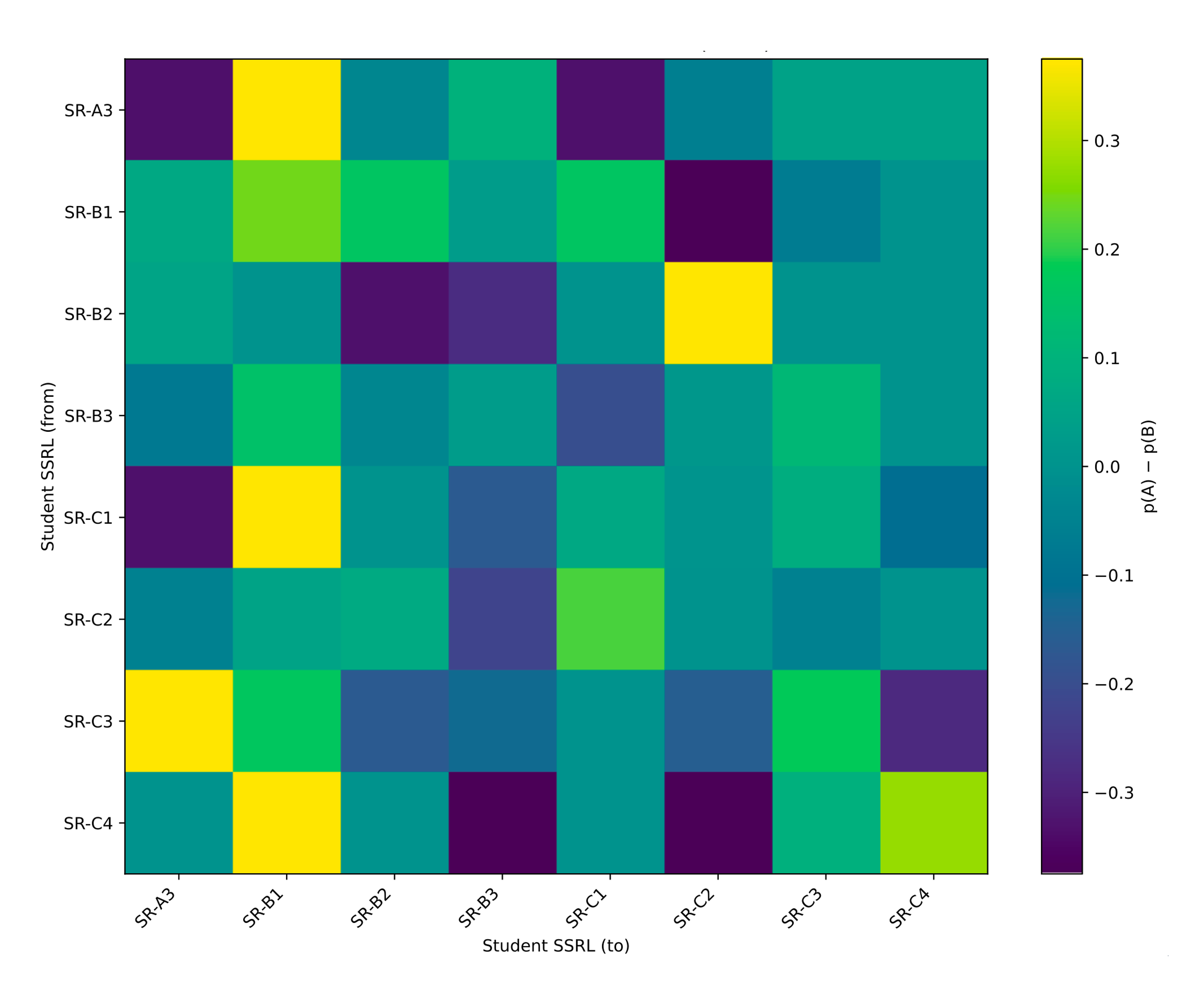}
    \caption{Transition heatmap of shared regulation (SSRL) behaviors across the two AI configurations. Each cell represents the normalized frequency of transitions from one SSRL category to another (from → to).}
    \label{fig:findings 4}
\end{figure*}

The shared-AI configuration fostered more synchronized, group-centered regulation cycles. Transitions often moved from socio-emotional or cognitive alignment toward coordinated action. In particular, Condition A showed stronger pathways such as SR-A3 → SR-B1 (climate regulation → coordinated action). In these moments, students often adjusted group atmosphere or relational stance.  (“It even uses emojis—haha, that’s so cute! I think the idea sounds good. Maybe we can try it this way.”) before collectively taking action. Similarly, transitions like SR-C1 → SR-B1 (goal alignment → coordinated action) and SR-C4 → SR-B1 (creative ideation → coordinated action) appeared more frequently in Condition A, suggesting that shared access to AI encouraged students to establish shared understanding or refine ideas collectively before taking action. In addition, Condition A also demonstrated stronger transitions involving collective interpretation and shared meaning reconstruction, especially SR-B2 → SR-C2 (progress monitoring → collective understanding repair). When encountering difficulties, students in Condition A were more likely to pause and clarify there understanding together (“What did you mean when you said the monitoring device would be hard to build?") before proceeding. This pattern reflects a more collective regulatory process, where monitoring prompts shared group-level cognitive work.

In contrast, the individual-AI configuration resulted in more fragmented or tool-centered regulation patterns. Transitions such as SR-C3 → SR-C4 (strategy discussion → idea elaboration) were more pronounced in Condition B, indicating that suggesting that students often developed their ideas independently—guided by their own AI agents—before attempting to align them as a group. Likewise, SR-C4 → SR-B3 (idea elaboration → tool-use negotiation) occurred more often in Condition B, reflecting a tendency to negotiate the use of individual AI tools rather than move directly into coordinated group action. These patterns suggest that individual AI access distributed regulatory effort across students, making alignment more likely to occur after independent work rather than synchronized, shared progression.

Taken together, these findings show that the shared-AI setup supported cohesive, sequentially aligned group regulation, where students jointly shaped the group climate, clarified goals, monitored challenges, and moved into coordinated action. In contrast, the individual-AI setup encouraged parallel, tool-mediated, and partially fragmented regulation. Students often worked through ideas separately with their own AI agents, which later required additional effort to reconcile differences and restore group alignment These reults highlights that AI configuration not only affects who interacts with AI, but also how groups regulate, synchronize, and coordinate their collaborative work.

\section{Discussion}

\subsection{AI Configuration as a Structural Condition for Collaborative Regulation}

Our study show that AI configuration does not simply affect interaction frequency; rather it alters the structural conditions under which collaboration unfolds. Shared AI access was associated with convergence-oriented collective reasoning, whereas individual AI access was linked to more distributed and ,at times, fragmented regulation cycles. Importantly, these patterns suggest that AI configuration acts as a macro-level design variable that shapes how regulation, alignment, and teacher orchestration emerge in classroom systems.

These findings extend prior research on AI-supported collaborative learning, which has largely focused on functional affordances (e.g., feedback quality, prompt design) rather than access structure~\cite{lipnevich2021review,zha2025mentigo,ba2025investigating}. While previous studies have suggested that shared artifacts promote common ground formation~\cite{moreillon2015increasing}, our results demonstrate that AI itself can assume a similar role when access is shared. Conversely, consistent with research on distributed cognition, individualized AI appears to introduce multiply epistemic centers, thereby increasing divergence and the need for coordination and reconciliation~\cite{gupta2025fostering}. Thus, these findings shift the focus from whether AI enhances collaboration to the conditions under which specific access configurations stabilize or destabilize collective regulation.

\subsection{From Shared Anchoring to Distributed Repair: Two Regulatory Regimes}

In this study, distinct regulatory patterns emerged under different AI configurations. Shared AI access was associaited with convergence-oriented regulation, as students aligned interpretations, stabilized group climate, and coordinated action around common AI outputs. Individual AI access, by contrast, generated more distributed trajectories that required later reconciliation, often triggered by discrepancies across multiple AI responses. These findings indicate that AI configuration influences whether collaborative regulation develops through shared anchoring or through distributed repair.

This distinction aligns with prior CSCL research emphasizing the importance of shared artifacts in establishing common ground~\cite{suthers2003representational}. Shared representations have long been shown to stabilize collective cognition by reducing interpretive divergence. Our findings extend this perspective by demonstrating that AI itself can function as such a shared epistemic anchor. At the same time, research on distributed cognition suggests that decentralization increases interpretive diversity and coordination demands. The individual-AI condition in our study reflects this dynamic, where multiple AI agents introduce parallel epistemic streams that required subsequent integration.

Importantly, our results nuance existing assumptions about personalization. While individualized AI is often framed as enhancing autonomy and exploration, the present findings suggest that it may also increase divergence and redirect regulatory effort toward repair and coordination. Rather than viewing personalization as inherently beneficial, our study highlights its structural implications for collective regulation. In doing so, we contribute a configuration-sensitive perspective to AI-supported collaborative learning research.

\subsection{Redistribution of Teacher Orchestration}

The two regulatory regimes were accompanied by distinct patterns of teacher orchestration. When AI functioned as a shared anchor, teachers primarily engaged in proactive, whole-class guidance. Because students worked from common AI outputs, clarification processes were often handled collectively, reducing the need for frequent individualized intervention. In contrast, individualized AI access was associated with micro-level teacher involvement, particularly when students encountered divergent outputs requiring interpretation or reconciliation.

This pattern resonates with classroom orchestration literature, which argues that technology does not eliminate instructional complexity but redistributes it \cite{dillenbourg2013design,song2021review,roschelle2013classroom}. Although AI personalization is often assumed to reduce teacher workload, empirical evidence on orchestration effects remains limited. Our findings suggest that individualized AI may increase coordination demands by multiplying epistemic entry points within a group. In such settings, teachers may need to manage heterogeneity across parallel reasoning paths rather than address a shared misunderstanding.

Thus, AI configuration reshapes the ecology of orchestration. Shared AI centralizes interpretive authority, enabling teachers to guide collective progression more directly. Individual AI decentralizes epistemic control, positioning teachers as mediators across divergent cognitive trajectories. This insight advances prior work by linking AI configuration to shifts in teacher regulatory positioning rather than merely to changes in interaction frequency.

\subsection{Implications for AI-Supported CSCL Design}

These findings carry implications for the design of AI-supported collaborative learning environments. When the pedagogical objective emphasizes alignment, shared conceptual grounding, and coordinated reasoning, shared AI configurations may provide structural advantages. By anchoring discussion around common outputs,shared access can reduce fragmentation and promotes stable regulatory loops. In contrast, if the objective prioritizes divergent exploration, ideation diversity, or individualized experimentation, distributed AI configurations may be appropriate. However, instructional designers should anticipate increased reconciliation demands and provide scaffolds for convergence \cite{keulemans2025lightening}. Hybrid approaches—such as staged transitions from shared to individualized AI—may help balance collective anchoring with exploratory breadth\cite{liu2026generative}.

Prior research on CSCL design highlights the importance of structuring interaction scripts and representational tools to support regulation\cite{suthers2003representational,feng2023effectiveness}. Our study suggests that AI configuration itself functions as a macro-level script that shape regulatory flow. Future research should therefore examine how configuration interacts with task complexity, learner expertise, and temporal sequencing. Understanding AI as part of a broader regulatory ecology may enable more intentional and pedagogically aligned integration.

\section{Limitations and future research}

Several limitations should be acknowledged. Although the study was conducted in an authentic classroom setting, it was situated within a single instructional context and school environment. The observed interaction patterns may therefore partially reflect local pedagogical culture or task design. Future research should examine whether the configuration-sensitive regulatory regimes identified here generalize across subject domains, age groups, and longer-term collaborative learning projects. In addition, our analyses relied on coded discourse data combined with LSA and ONA modelling. While these approaches capture structural coupling and transition dynamics, they infer regulation from observable interaction rather than directly measuring learners’ internal cognitive or motivational processes. The granularity of coding and modelling parameters may also influence the sensitivity of detected network structures. Integrating multimodal data sources or validated self-report measures could strengthen interpretive validity. The intervention was further bounded within a relatively short time frame. Students’ strategies for engaging with shared versus individual AI agents may evolve with increased familiarity. Longitudinal studies are therefore needed to examine how configuration effects stabilize or transform over time and whether hybrid or staged AI configurations can support adaptive shifts between collective alignment and distributed exploration. Finally, although our findings suggest a redistribution of teacher orchestration across configurations, we did not directly measure teacher workload or cognitive load. Future work should investigate how different AI arrangements reshape orchestration demands and explore teacher-facing analytic supports to sustain AI-mediated collaborative learning at scale.

\section{Conclusion}
In this paper, we show that AI configuration is not a neutral technical choice but a structural condition that reshapes the regulatory ecology of classroom collaboration. Drawing on discourse coding, LSA, and network modelling within an authentic CSCL classroom, we demonstrate that shared and individual AI configurations produce systematically different patterns of interaction distribution, AI–student coupling, shared regulation, and teacher orchestration. Shared AI access functioned as a collective cognitive anchor, fostering convergence, deeper cognitive progression, and inspired group-level regulation. Individual AI access, by contrast, generated more distributed and repair-driven interaction cycles, increasing evaluative and coordination demands and redistributing orchestration work toward the teacher. By situating generative AI within a real classroom setting rather than isolated laboratory tasks, this study contributes empirical evidence to ongoing debates about how AI integrates into collaborative learning environments. More broadly, our findings suggest that the pedagogical impact of AI lies not only in what the system can generate, but in how its access structure organizes participation, regulation, and alignment within the learning community. Designing AI-supported CSCL, therefore, requires attention to configuration as a pedagogical variable in its own right.

\bibliography{sample}

@article{keulemans2025lightening,
  title={Lightening the load: Unveiling the factors influencing teacher orchestration load in synchronous hybrid education},
  author={Keulemans, Tine and Depaepe, Fien and Raes, Annelies},
  journal={Learning and Instruction},
  volume={99},
  pages={102189},
  year={2025},
  publisher={Elsevier}
}

@article{feng2023effectiveness,
  title={Effectiveness of the functions of classroom orchestration systems: A systematic review and meta-analysis},
  author={Feng, Shuo and Zhang, Lishan and Wang, Shuwen and Cai, Zhihui},
  journal={Computers \& Education},
  volume={203},
  pages={104864},
  year={2023},
  publisher={Elsevier}
}

@article{roschelle2013classroom,
  title={Classroom orchestration: synthesis},
  author={Roschelle, Jeremy and Dimitriadis, Yannis and Hoppe, Ulrich},
  journal={Computers \& Education},
  volume={69},
  pages={523--526},
  year={2013},
  publisher={Elsevier}
}

@article{song2021review,
  title={A review of how class orchestration with technology has been conducted for pedagogical practices},
  author={Song, Yanjie},
  journal={Educational Technology Research and Development},
  volume={69},
  number={3},
  pages={1477--1503},
  year={2021},
  publisher={Springer}
}

@article{dillenbourg2013design,
  title={Design for classroom orchestration},
  author={Dillenbourg, Pierre},
  journal={Computers \& education},
  volume={69},
  pages={485--492},
  year={2013},
  publisher={Elsevier}
}

@inproceedings{han2024teams,
  title={When teams embrace AI: human collaboration strategies in generative prompting in a creative design task},
  author={Han, Yuanning and Qiu, Ziyi and Cheng, Jiale and Lc, Ray},
  booktitle={Proceedings of the 2024 CHI Conference on Human Factors in Computing Systems},
  pages={1--14},
  year={2024}
}

@article{law2025role,
  title={The role of generative AI in collaborative problem-solving of authentic challenges},
  author={Law, Nancy and Wang, Nan and Ma, Ming and Liu, Zhichun and Lei, Leon and Feng, Shuhui and Hu, Xiao and Tsao, Jack},
  journal={British Journal of Educational Technology},
  year={2025},
  publisher={Wiley Online Library}
}

@article{teng2024chatgpt,
  title={“ChatGPT is the companion, not enemies”: EFL learners’ perceptions and experiences in using ChatGPT for feedback in writing},
  author={Teng, Mark Feng},
  journal={Computers and Education: Artificial Intelligence},
  volume={7},
  pages={100270},
  year={2024},
  publisher={Elsevier}
}

@article{sharples2023towards,
  title={Towards social generative AI for education: theory, practices and ethics},
  author={Sharples, Mike},
  journal={Learning: Research and Practice},
  volume={9},
  number={2},
  pages={159--167},
  year={2023},
  publisher={Taylor \& Francis}
}

@incollection{salomon1992effects,
  title={Effects with and of computers and the study of computer-based learning environments},
  author={Salomon, Gavriel},
  booktitle={Computer-based learning environments and problem solving},
  pages={249--263},
  year={1992},
  publisher={Springer}
}

@article{li2022influence,
  title={The influence of socially shared regulation on computational thinking performance in cooperative learning},
  author={Li, Jiansheng and Liu, Jiao and Yuan, Rui and Shadiev, Rustam},
  journal={Educational Technology \& Society},
  volume={25},
  number={1},
  pages={48--60},
  year={2022},
  publisher={JSTOR}
}

@article{treffinger1995creative,
  title={Creative problem solving: Overview and educational implications},
  author={Treffinger, Donald J},
  journal={Educational psychology review},
  volume={7},
  number={3},
  pages={301--312},
  year={1995},
  publisher={Springer}
}

@article{hadwin2011self,
  title={Self-regulated, co-regulated, and socially shared regulation of learning},
  author={Hadwin, Allyson Fiona and J{\"a}rvel{\"a}, Sanna and Miller, Mariel},
  journal={Handbook of self-regulation of learning and performance},
  volume={30},
  pages={65--84},
  year={2011}
}

@article{sayeed2025collaclassroom,
  title={CollaClassroom: An AI-Augmented Collaborative Learning Platform with LLM Support in the Context of Bangladeshi University Students},
  author={Sayeed, Salman and Saiem, Bijoy Ahmed and Sany, Al-Amin and Sharmin, Sadia and Islam, ABM},
  journal={arXiv preprint arXiv:2511.11823},
  year={2025}
}

@article{daryanto2026human,
  title={Human-Human-AI Triadic Programming: Uncovering the Role of AI Agent and the Value of Human Partner in Collaborative Learning},
  author={Daryanto, Taufiq and Ding, Xiaohan and Ping, Kaike and Wilhelm, Lance T and Chen, Yan and Brown, Chris and Rho, Eugenia H},
  journal={arXiv preprint arXiv:2601.12134},
  year={2026}
}

@article{feng2025group,
  title={Group interaction patterns in generative AI-supported collaborative problem solving: Network analysis of the interactions among students and a GAI chatbot},
  author={Feng, Shihui},
  journal={British Journal of Educational Technology},
  year={2025},
  publisher={Wiley Online Library}
}

@article{suthers2006technology,
  title={Technology affordances for intersubjective meaning making: A research agenda for CSCL},
  author={Suthers, Daniel D},
  journal={International Journal of Computer-supported collaborative learning},
  volume={1},
  number={3},
  pages={315--337},
  year={2006},
  publisher={Springer}
}

@inproceedings{roschelle1995construction,
  title={The construction of shared knowledge in collaborative problem solving},
  author={Roschelle, Jeremy and Teasley, Stephanie D},
  booktitle={Computer supported collaborative learning},
  pages={69--97},
  year={1995},
  organization={Springer}
}

@article{yan2024promises,
  title={Promises and challenges of generative artificial intelligence for human learning},
  author={Yan, Lixiang and Greiff, Samuel and Teuber, Ziwen and Ga{\v{s}}evi{\'c}, Dragan},
  journal={Nature Human Behaviour},
  volume={8},
  number={10},
  pages={1839--1850},
  year={2024},
  publisher={Nature Publishing Group UK London}
}

@article{zheng2025generative,
  title={A generative artificial intelligence-enhanced multiagent approach to empowering collaborative problem solving across different learning domains},
  author={Zheng, Lanqin and Shi, Zhe and Gao, Lei},
  journal={Computers \& Education},
  pages={105489},
  year={2025},
  publisher={Elsevier}
}

@article{chen2025meetmap,
  title={MeetMap: Real-Time Collaborative Dialogue Mapping with LLMs in Online Meetings},
  author={Chen, Xinyue and Yap, Nathan and Lu, Xinyi and Gunal, Aylin and Wang, Xu},
  journal={Proceedings of the ACM on Human-Computer Interaction},
  volume={9},
  number={2},
  pages={1--35},
  year={2025},
  publisher={ACM New York, NY, USA}
}

@inproceedings{zhang2025ladica,
  title={LADICA: a large shared display interface for generative AI cognitive assistance in co-located team collaboration},
  author={Zhang, Zheng and Peng, Weirui and Chen, Xinyue and Cao, Luke and Li, Toby Jia-Jun},
  booktitle={Proceedings of the 2025 CHI Conference on Human Factors in Computing Systems},
  pages={1--22},
  year={2025}
}

@inproceedings{davis2025co,
  title={The Co-Creative Design Framework for Hybrid Intelligence},
  author={Davis, Nicholas and Sherson, Jacob and Rafner, Janet},
  booktitle={Proceedings of the 2025 Conference on Creativity and Cognition},
  pages={560--572},
  year={2025}
}

@incollection{stahl2013theories,
  title={Theories of collaborative cognition: Foundations for CSCL and CSCW together},
  author={Stahl, Gerry},
  booktitle={Computer-Supported Collaborative Learning at the Workplace: CSCL@ Work},
  pages={43--63},
  year={2013},
  publisher={Springer}
}

@article{malmberg2015promoting,
  title={Promoting socially shared regulation of learning in CSCL: Progress of socially shared regulation among high-and low-performing groups},
  author={Malmberg, Jonna and J{\"a}rvel{\"a}, Sanna and J{\"a}rvenoja, Hanna and Panadero, Ernesto},
  journal={Computers in human behavior},
  volume={52},
  pages={562--572},
  year={2015},
  publisher={Elsevier}
}

@article{jarvela2015enhancing,
  title={Enhancing socially shared regulation in collaborative learning groups: Designing for CSCL regulation tools},
  author={J{\"a}rvel{\"a}, Sanna and Kirschner, Paul A and Panadero, Ernesto and Malmberg, Jonna and Phielix, Chris and Jaspers, Jos and Koivuniemi, Marika and J{\"a}rvenoja, Hanna},
  journal={Educational Technology Research and Development},
  volume={63},
  number={1},
  pages={125--142},
  year={2015},
  publisher={Springer}
}

@inproceedings{chen2025analyzing,
  title={Analyzing Students' Use of Generative Artificial Intelligence in Collaborative Problem Solving},
  author={Chen, Xiuyu and Feng, Shihui},
  booktitle={Companion Publication of the 2025 Conference on Computer-Supported Cooperative Work and Social Computing},
  pages={385--389},
  year={2025}
}

@article{perifanou2025collaborative,
  title={Collaborative uses of GenAI tools in project-based learning},
  author={Perifanou, Maria and Economides, Anastasios A},
  journal={Education Sciences},
  volume={15},
  number={3},
  pages={354},
  year={2025},
  publisher={MDPI}
}

@article{ogunleye2024systematic,
  title={A systematic review of generative AI for teaching and learning practice},
  author={Ogunleye, Bayode and Zakariyyah, Kudirat Ibilola and Ajao, Oluwaseun and Olayinka, Olakunle and Sharma, Hemlata},
  journal={Education Sciences},
  volume={14},
  number={6},
  pages={636},
  year={2024},
  publisher={MDPI}
}

@article{liu2025integrating,
  title={Integrating generative Artificial Intelligence into student learning: A systematic review from a TPACK perspective},
  author={Liu, Xiaofan and Zhong, Baichang},
  journal={Educational Research Review},
  pages={100741},
  year={2025},
  publisher={Elsevier}
}

@article{guoa2025student,
  title={Student-AI Collaborative Creative Problem-Solving: The Role of Human Agency},
  author={Guoa, Wenxin and Lianga, Zheng and Wang, Chengcheng and Li, Xing and Hu, Huiqing and Chen, Shi and Yua, Quanlei and Zhaoa, Qingbai},
  journal={Computers \& Education},
  pages={105433},
  year={2025},
  publisher={Elsevier}
}

@article{cukurova2025interplay,
  title={The interplay of learning, analytics and artificial intelligence in education: A vision for hybrid intelligence},
  author={Cukurova, Mutlu},
  journal={British Journal of Educational Technology},
  volume={56},
  number={2},
  pages={469--488},
  year={2025},
  publisher={Wiley Online Library}
}

@article{belkina2025implementing,
  title={Implementing generative AI (GenAI) in higher education: A systematic review of case studies},
  author={Belkina, Marina and Daniel, Scott and Nikolic, Sasha and Haque, Rezwanul and Lyden, Sarah and Neal, Peter and Grundy, Sarah and Hassan, Ghulam M},
  journal={Computers and Education: Artificial Intelligence},
  pages={100407},
  year={2025},
  publisher={Elsevier}
}

@article{wei2025effects,
  title={The effects of generative AI on collaborative problem-solving and team creativity performance in digital story creation: an experimental study},
  author={Wei, Xiaodong and Wang, Lei and Lee, Lap-Kei and Liu, Ruixue},
  journal={International Journal of Educational Technology in Higher Education},
  volume={22},
  number={1},
  pages={23},
  year={2025},
  publisher={Springer}
}

@article{zhuang2025enhancing,
  title={Enhancing language learning through generative AI feedback on picture-cued writing tasks},
  author={Zhuang, Yipeng and Zhao, Ruibin and Xie, ZhiWei and Yu, Philip LH},
  journal={Computers and Education: Artificial Intelligence},
  pages={100450},
  year={2025},
  publisher={Elsevier}
}

@article{kovari2025systematic,
  title={A systematic review of AI-powered collaborative learning in higher education: Trends and outcomes from the last decade},
  author={Kovari, Attila},
  journal={Social Sciences \& Humanities Open},
  volume={11},
  pages={101335},
  year={2025},
  publisher={Elsevier}
}

@article{liu2026generative,
  title={Generative AI and human mentorship in creative problem solving},
  author={Liu, Weijing and Zha, Siyu and Xu, Yingqing},
  journal={Educational Technology \& Society},
  volume={29},
  number={1},
  pages={332--358},
  year={2026},
  publisher={JSTOR}
}

@inproceedings{lipnevich2021review,
  title={A review of feedback models and theories: Descriptions, definitions, and conclusions},
  author={Lipnevich, Anastasiya A and Panadero, Ernesto},
  booktitle={Frontiers in Education},
  volume={6},
  pages={720195},
  year={2021},
  organization={Frontiers}
}

@inproceedings{zha2025mentigo,
  title={Mentigo: An Intelligent Agent for Mentoring Students in the Creative Problem Solving Process},
  author={Zha, Siyu and Liu, Yujia and Zheng, Chengbo and Xu, Jiaqi and Yu, Fuze and Gong, Jiangtao and Xu, Yingqing},
  booktitle={Proceedings of the 2025 CHI Conference on Human Factors in Computing Systems},
  pages={1--22},
  year={2025}
}

@article{moreillon2015increasing,
  title={Increasing interactivity in the online learning environment: Using digital tools to support students in socially constructed meaning-making},
  author={Moreillon, Judi},
  journal={TechTrends},
  volume={59},
  number={3},
  pages={41--47},
  year={2015},
  publisher={Springer}
}

@article{gupta2025fostering,
  title={Fostering collective intelligence in human--AI collaboration: laying the groundwork for COHUMAIN},
  author={Gupta, Pranav and Nguyen, Thuy Ngoc and Gonzalez, Cleotilde and Woolley, Anita Williams},
  journal={Topics in cognitive science},
  volume={17},
  number={2},
  pages={189--216},
  year={2025},
  publisher={Wiley Online Library}
}

@article{giannakos2025promise,
  title={The promise and challenges of generative AI in education},
  author={Giannakos, Michail and Azevedo, Roger and Brusilovsky, Peter and Cukurova, Mutlu and Dimitriadis, Yannis and Hernandez-Leo, Davinia and J{\"a}rvel{\"a}, Sanna and Mavrikis, Manolis and Rienties, Bart},
  journal={Behaviour \& Information Technology},
  volume={44},
  number={11},
  pages={2518--2544},
  year={2025},
  publisher={Taylor \& Francis}
}

@article{zercher2025can,
  title={How Can Teams Benefit From AI Team Members? Exploring the Effect of Generative AI on Decision-Making Processes and Decision Quality in Team--AI Collaboration},
  author={Zercher, D{\'e}sir{\'e}e and Jussupow, Ekaterina and Benke, Ivo and Heinzl, Armin},
  journal={Journal of Organizational Behavior},
  year={2025},
  publisher={Wiley Online Library}
}

@article{kim2025socially,
  title={Socially shared regulation of learning and artificial intelligence: Opportunities to support socially shared regulation},
  author={Kim, Jinhee and Detrick, Rita and Yu, Seongryeong and Song, Yukyeong and Bol, Linda and Li, Na},
  journal={Education and Information Technologies},
  pages={1--39},
  year={2025},
  publisher={Springer}
}

@article{ba2025investigating,
  title={Investigating the impact of ChatGPT-assisted feedback on the dynamics and outcomes of online inquiry-based discussion},
  author={Ba, Shen and Zhan, Ying and Huang, Lingyun and Lu, Guoqing},
  journal={British Journal of Educational Technology},
  year={2025},
  publisher={Wiley Online Library}
}

@incollection{suthers2003representational,
  title={Representational guidance for collaborative inquiry},
  author={Suthers, Daniel D},
  booktitle={Arguing to learn: Confronting cognitions in computer-supported collaborative learning environments},
  pages={27--46},
  year={2003},
  publisher={Springer}
}

@String{Computing = "Computing" }

@String{Computer = "{IEEE} Computer" }

@String{Springer = "Springer-Verlag" }

@ArtifactSoftware{R,
    title = {R: A Language and Environment for Statistical Computing},
    author = {{R Core Team}},
    organization = {R Foundation for Statistical Computing},
    address = {Vienna, Austria},
    year = {2019},
    url = {https://www.R-project.org/},
}

@article{zha2025designing,
  title={Designing child-centric AI learning environments: Insights from an LLM-powered creative project-based learning study},
  author={Zha, Siyu and Qiao, Yuehan and Hu, Qingyu and Li, Zhonghseng and Gong, Jiangtao and Xu, Yingqing},
  journal={International Journal of Human-Computer Studies},
  pages={103602},
  year={2025},
  publisher={Elsevier}
}

@inproceedings{liu2025bricksmart,
  title={BrickSmart: Leveraging Generative AI to Support Children's Spatial Language Learning in Family Block Play},
  author={Liu, Yujia and Zha, Siyu and Zhang, Yuewen and Wang, Yanjin and Zhang, Yangming and Xin, Qi and Nie, Lun Yiu and Zhang, Chao and Xu, Yingqing},
  booktitle={Proceedings of the 2025 CHI Conference on Human Factors in Computing Systems},
  pages={1--19},
  year={2025}
}






\clearpage
\section{Appendix}

\begin{longtable}{p{3cm} p{2cm} p{3.5cm}p{4cm}}
\caption{Full coding framework for human--AI--teacher interaction analysis}
\label{tab:full_coding_scheme}\\

\toprule
\textbf{Category} & \textbf{Code} & \textbf{Label}  & \textbf{Operational definition} \\
\midrule
\endfirsthead

\toprule
\textbf{Category} & \textbf{Code} & \textbf{Label} & \textbf{Operational definition} \\
\midrule
\endhead

\midrule
\multicolumn{3}{r}{\footnotesize Continued on next page} \\
\endfoot

\bottomrule
\endlastfoot

\textbf{Student Orientation} \\
& S-S & Student--student interaction & Turn directed to one or more peers, involving exchange, coordination, or discussion among students. \\
& S-AI & Student--AI interaction & Turn directed to the AI system, including querying, responding to, or interacting with AI output. \\
& S-T & Student--teacher interaction & Turn directed to the teacher, including asking for clarification or responding to teacher input. \\
& S-Screen & Screen operation & Non-verbal or minimal-verbal interaction focused on device operation (e.g., typing, scrolling, navigating interface). \\
\midrule

\textbf{Student Affect} \\
& A1 & Positive affect & Expression of positive emotional tone (e.g., enthusiasm, interest, satisfaction, encouragement). \\
& A2 & Neutral affect & Emotionally neutral expression without clear positive or negative valence. \\
& A3 & Negative affect & Expression of frustration, confusion, dissatisfaction, or discouragement. \\
\midrule

\textbf{Student Cognition} \\
& C1 & Comprehension & Statements demonstrating understanding, recall, or clarification of task-related information. \\
& C2 & Implementation and analysis & Applying ideas to the task, examining feasibility, identifying constraints, or analyzing implications. \\
& C3 & Evaluation & Judging the quality, feasibility, or appropriateness of ideas, plans, or outputs. \\
& C4 & Creation & Generating novel ideas, proposing new plans, or synthesizing elements into a new solution. \\
& C5 & Non-task-related & Utterances unrelated to the learning task or collaborative goal. \\
\midrule

\textbf{Student Behavior} \\
& B1 & Initiating questions & Student introduces a new question that advances or redirects discussion. \\
& B2 & Follow-up inquiry & Student asks a clarifying or extending question building on prior discourse. \\
& B3 & Idea or state reporting & Reporting personal ideas, status updates, or observations without directly querying others. \\
& B4 & Responding to AI & Verbal reaction to AI output, including agreement, disagreement, or interpretation. \\
& B5 & Probing or testing AI & Challenging, refining, or experimentally modifying AI input to explore its response. \\
& B6 & Copying AI output & Directly adopting or repeating AI-generated content with minimal modification. \\
& B7 & Revising or extending AI output & Modifying, elaborating, or integrating AI output into new student-generated contributions. \\
& B8 & Evaluating AI output & Critically assessing the quality, relevance, or feasibility of AI-generated content. \\
\midrule

\textbf{Shared Regulation -- Cognitive} \\
& SR-C1 & Goal setting and alignment & Joint articulation or clarification of shared goals, objectives, or task direction. \\
& SR-C2 & Collective understanding & Negotiating or clarifying shared interpretation of task content or AI output. \\
& SR-C3 & Learning strategy discussion & Discussion of strategies, approaches, or decision-making processes for task completion. \\
& SR-C4 & Metacognitive reflection & Collective reflection on progress, effectiveness, or need to adjust strategies. \\
\midrule

\textbf{Shared Regulation -- Behavioral} \\
& SR-B1 & Coordinated action & Explicit coordination of roles, responsibilities, or next actions among group members. \\
& SR-B2 & Progress monitoring & Joint tracking or reporting of task progress or completion status. \\
& SR-B3 & Tool use negotiation & Discussion or negotiation about when and how to use AI or other tools. \\
\midrule

\textbf{Shared Regulation -- Affective} \\
& SR-A1 & Emotional support & Providing encouragement, reassurance, or emotional validation within the group. \\
& SR-A2 & Motivation maintenance & Statements aimed at sustaining group effort or persistence. \\
& SR-A3 & Climate regulation & Managing group atmosphere, tension, humor, or relational dynamics. \\
\midrule

\textbf{AI Functions} \\
& F1 & Information provision & Providing factual information, data, or task-relevant content without extended explanation. \\
& F2 & Conceptual explanation & Explaining underlying concepts, principles, or reasoning processes. \\
& F3 & Illustrative example & Providing concrete examples to clarify or demonstrate a concept or idea. \\
& F4 & Idea or solution generation & Generating new ideas, suggestions, or possible solutions. \\
& F5 & Evaluation and feedback & Offering evaluative comments, critiques, or improvement suggestions. \\
& F6 & Question prompting & Asking guiding or probing questions intended to stimulate thinking. \\
& F7 & Affective encouragement  & Expressing supportive or motivational language toward the learner(s). \\
& F8 & Noisy or irrelevant output & Providing output unrelated, misleading, or inconsistent with the task. \\
\midrule

\textbf{Teacher Regulation -- Cognitive} \\
& TR-C1 & Goal clarification & Clarifying learning objectives, task requirements, or conceptual expectations. \\
& TR-C2 & Evaluation & Assessing the quality, feasibility, or correctness of student work. \\
\midrule

\textbf{Teacher Regulation -- Behavioral} \\
& TR-B1 & Pacing control & Managing time, sequencing, or workflow during the task. \\
& TR-B2 & Instructional guidance & Providing procedural or strategic guidance for task progression. \\
& TR-B3 & Tool guidance & Directing or advising students on the use of AI or other tools. \\
\midrule

\textbf{Teacher Regulation -- Affective} \\
& TR-A1 & Emotional support & Providing encouragement, reassurance, or affective affirmation. \\
\midrule

\textbf{Teacher Behavior} \\
& TB1 & Teacher-initiated instructional guidance & Teacher proactively introduces new directions, redirects focus, or initiates instructional moves. \\
& TB2 & Teacher responses to student-initiated questions & Teacher reacts to student inquiries by clarifying, explaining, or providing additional information. \\

\end{longtable}

\begin{longtable}{p{4cm} p{4cm} r r}
\caption{Distribution of interaction behaviors across coding groups (Condition A and Condition B)}\\
\toprule
\textbf{Coding Group} & \textbf{Behavior} & \textbf{Count-A} & \textbf{Count-B}\\
\midrule
\endfirsthead

\toprule
\textbf{Coding Group} & \textbf{Behavior} & \textbf{Count-A} & \textbf{Count-B}\\
\midrule
\endhead

\midrule
\multicolumn{4}{r}{\footnotesize Continued on next page} \\
\endfoot

\bottomrule
\endlastfoot

\textbf{Role} \\
\quad S & Student & 304 & 363\\
\quad T & Teacher & 29 & 40\\
\quad AI & AI & 78 & 126\\
\addlinespace[4pt]

\textbf{S-I (Student Interaction)} \\
\quad S-S & Student--Student & 224 & 194\\
\quad S-AI & Student--AI & 74 & 150\\
\quad S-T & Student--Teacher & 0 & 16\\
\quad S-Screen & Student--Screen & 5 & 4\\
\addlinespace[4pt]

\textbf{AI (AI Functions)} \\
\quad F1 & Information provision & 24 & 64 \\
\quad F2 & Conceptual explanation & 9 & 3 \\
\quad F3 & Illustrative example & 22 & 49 \\
\quad F4 & Idea/solution generation & 23 & 17 \\
\quad F5 & Evaluation/feedback & 18 & 35 \\
\quad F6 & Question prompting & 44 & 43 \\
\quad F7 & Affective encouragement & 65 & 103 \\
\addlinespace[4pt]

\textbf{S-Affect (Student Affect)} \\
\quad A1 & Positive & 11 & 8 \\
\quad A2 & Neutral & 282 & 332 \\
\quad A3 & Negative & 10 & 22 \\
\addlinespace[4pt]

\textbf{S-Cognition (Student Cognition)} \\
\quad C1 & Comprehension & 67 & 75 \\
\quad C2 & Implementation & 70 & 40 \\
\quad C3 & Evaluation & 7 & 4 \\
\quad C4 & Creation & 23 & 13 \\
\quad C5 & Non-task-related & 136 & 234 \\
\addlinespace[4pt]

\textbf{S-Behavior (Student Behavior)} \\
\quad B1 & Initiating questions & 62 & 56 \\
\quad B2 & Follow-up inquiry & 10 & 6 \\
\quad B3 & Idea/state reporting & 124 & 121 \\
\quad B4 & Responding to AI & 12 & 29 \\
\quad B5 & Probing/testing AI & 20 & 27 \\
\quad B6 & Copying AI output & 9 & 6 \\
\quad B7 & Revising/extending AI output & 4 & 9 \\
\quad B8 & Evaluating AI output & 18 & 25 \\
\addlinespace[4pt]

\textbf{SR-A / SR-B / SR-C (Shared Regulation)} \\
\quad SR-A1 & Emotional support & 2 & 5 \\
\quad SR-A2 & Motivation maintenance & 3 & 0 \\
\quad SR-A3 & Climate regulation & 27 & 34 \\
\quad SR-B1 & Coordinated action & 44 & 22 \\
\quad SR-B2 & Progress monitoring & 11 & 17 \\
\quad SR-B3 & Tool use negotiation & 28 & 49 \\
\quad SR-C1 & Goal setting & 15 & 6 \\
\quad SR-C2 & Collective understanding & 69 & 76 \\
\quad SR-C3 & Strategy discussion & 19 & 30 \\
\quad SR-C4 & Metacognitive reflection & 12 & 10 \\
\addlinespace[4pt]

\textbf{TR (Teacher Regulation)} \\
\quad TR-C1 & Goal clarification & 17 & 8 \\
\quad TR-C2 & Evaluation & 0 & 6 \\
\quad TR-B1 & Pacing control & 15 & 8 \\
\quad TR-B2 & Instructional guidance & 4 & 2 \\
\quad TR-B3 & Tool guidance & 8 & 22 \\
\quad TR-A1 & Emotional support & 2 & 2 \\
\addlinespace[4pt]

\textbf{TB (Teacher Behaviour)} \\
\quad TB1 & Teacher-initiated & 27 & 20\\
\quad TB2 & Teacher-responsive & 3 & 19 \\
\addlinespace[4pt]

\end{longtable}

\end{document}